# QUANTITATIVE MRI MOLECULAR IMAGING IN THE EVALUATION OF EARLY POST-MORTEM CHANGES IN MUSCLES. A FEASIBILITY STUDY ON A PIG PHANTOM

Daniela SAPIENZA[1*], Alessio ASMUNDO[1], Salvatore SILIPIGNI[2], Ugo BARBARO[2], Antonella CINQUEGRANI[2], Francesca GRANATA[2], Valeria BARRESI[3], Patrizia GUALNIERA[1], Antonio BOTTARI[2], Michele GAETA[2]

1 Department of Biomedical Sciences, Dental and of Morphological and Functional Images, Section of Legal Medicine, University of Messina, Via Consolare Valeria 1, 98125 Messina, Italy.

2 Department of Biomedical Sciences, Dental and of Morphological and Functional Images, Section of Radiological Sciences, University of Messina, Italy.

3 Department of Human Pathology in Adulthood and Evolutive Age, University of Messina, Messina, Italy.

*Corresponding Author daniela.sapienza@unime.it mobile: +39 3382800106
 ORCID ID:0000-0002-3595-7086 address: AOU Policlinico "G. Martino", via Consolare Valeria n. 1 - Messina (Italy)

## ABSTRACT

Estimating early postmortem interval (EPI) is a difficult task in daily forensic activity due to limitations of accurate and reliable methods.
The aim of the present work is to describe a novel approach in the estimation of EPI based on quantitative magnetic resonance molecular imaging (qMRMI) using a pig phantom since post-mortem degradation of pig meat is similar to that of human muscles.
On a pig phantom maintained at 20° degree, using a 1.5 T MRI scanner we performed 10 scans (every 4 hours) monitoring apparent diffusion coefficient (ADC), fractional anisotropy (FA) magnetization transfer ration (MTR), tractography and susceptibility weighted changes in muscles until 36 hours after death. Cooling of the phantom during the experiment was recorded. Histology was also obtained.
Pearson's Test was carried out for statistical correlation. We found a significative statistical inverse correlation between ADC, FA, MT and PMI.
Our preliminary data shows that post-mortem qMRMI is a potential powerful tool in accurately determining EPI and is worth of further investigation.

**Key Points:**
-Early postmortem interval (EPI) is a difficult task in daily forensic activity
-Novel methods for time of death estimation are still of high demand as well as postmortem muscle protein degradation
- Modern imaging techniques (CT and MRI) are diagnostic tools of increasing importance for the forensic and legal medicine in post-mortem investigations
- The role of quantitative MR molecular imaging (qMRMI) has emerged in as a fundamental approach in analysis of the tissues such in muscles
- qMRMI is peculiar in order to explore the potential usefulness of this non-invasive novel approach in the evaluation of early post-mortem interval.

# INTRODUCTION

Determination of the postmortem interval (PMI) and early postmortem interval (EPI) is one of the most challenging and difficult task in daily forensic activity due to limitations of accurate and reliable methods.

Since its introduction 30 years back, the normogram method by Henßge has been established as the standard procedure of temperature-based death time determination in the early post-mortem period [1–3].

However the validity of the normogram method seems to be problematic, death time estimates – and particularly their 95%-confidence interval limits – have to be interpreted carefully since systematic overestimation of the post-mortem interval in bodies of high mass and large surface area must be taken into account [4].

Other alterations after death include the development and regression of rigor mortis, the progression of livor mortis and algor mortis [5]. Although these methods are employed to delimitate the PMI in everyday forensic work, there are still great inaccuracies and limitations in many cases. Therefore, novel methods for time of death estimation that comply with the requirements of practice are still of high demand. Recently new promising approaches have been tested to characterize changes within the EPI including laboratoristic analysis on postmortem biochemistry [6], changes in oxidant/antioxidant parameters (malondialdehyde MDA, nitric oxide NO, total thiol as well as the activity of glutathione reductase GR, glutathione S transferase, and catalase) [7], and postmortem muscle protein degradation [8–10].

Modern imaging techniques (CT and MRI) are diagnostic tools of increasing importance for the forensic and legal medicine in post-mortem investigations, as an adjunct or as an alternative to autopsy. Although post-mortem imaging has some well-known limitations, today pre-autopsy post-mortem CT (PMCT) and/or post-mortem MR (PMMR) are considered useful procedures in many forensic institutes world-wide [11, 12] with good performances for depicting cause of death, traumatic findings in corpses and in the definition of post-mortem interval [11, 13–15].

Although MRI is widely used in clinical medicine, its routine diffusion into forensic medicine has been limited [14] by the longer acquisition time in comparison with PMCT and higher cost. In spite of this, in the last decade, PMMR has been emerging as a powerful tool in forensic death investigations and has the ability to enhance autopsy and uncover otherwise undetectable findings.

On the other hand, until now, PMMR has been largely based on conventional "morphological" MR T1 and T2 weighted and short-tau inversion recovery (STIR) sequences, which produce magnitude images whose signal mainly depends on proton density (PD), T1 and T2 relaxation times of tissues.

In addition, such images allow only qualitative evaluation and are consequently strongly linked to the subjective judgment and variable experience of radiologists.

In the last years, the role of quantitative MR molecular imaging (qMRMI) has emerged in clinical practice as a fundamental approach for the detection and diagnosis of diseases. For example diffusion weighted imaging has become the fundamental tool for early diagnosis of cerebral ischemia, allowing to display changes due to ischemia as early as 20 minutes after the stroke, and it is also a powerful diagnostic weapon in cancer [16, 17].

qMRMI allows to obtain numerical data and generates images whose signal depends on bio-physical tissue properties other than PD, T1 and T2 relaxation times, e.g. diffusion weighted sequence measures mobility of water molecules due to random Brownian motion within a tissue voxel. Besides information on isotropic and anisotropic water diffusion, obtained from Diffusion Weighted Imaging (DWI) and Fractional Anisotropy (FA), status of the macromolecular compartment including protein, obtained from Magnetization Transfer (MT), and presence of microscopic amount of air in tissues due to early putrefaction, estimated with Susceptibility Weighted Imaging (SWI), can be obtained in a fast and reliable way. Molecular imaging is a quantitative and objective method based on numeric values, which is independent from experience of radiologists. The role of quantitative MR molecular imaging is raising in clinical practice [18].

The aim of our study has been to evaluate the early post-mortem changes in muscles using qMRMI on a pig phantom in order to explore the potential usefulness of this non-invasive novel approach in the early post-mortem interval. To the best of our knowledge, no previous paper has been published on this topic in the scientific literature.

MATERIAL AND METHODS

A pig hind leg weighting 24 kg was used as a phantom. The pig was slaughtered according to standard procedure for food chain, under animal physician control. The pig, bred in a farm, was healthy with a well-developed muscle mass. The choice of a pig as a phantom was based on the demonstration that postmortem muscle changes were similar in time and temperature to that observed in humans.

No permission by our Ethic Committee was necessary because the phantom was obtained from the commercial food chain.

Time of pig death was recorded. The specimen was placed in rooms maintained at a constant temperature of 20° Celsius with a 40% of humidity from the first minutes after death until the end of the experiment, monitoring them with an ambient temperature-humidity CE device.

The first MRI scan was performed 3 hours after death. The last scan was carried out 36 hours after the first one. A total of 10 scans were obtained with an interval of 4 hours between each scan.

All the sequences were centered on the same axial axis of the pig leg localized using a superficial marker (adhesive plastic tape). The axial thickness of the muscles included into MRI scan was 15 cm (**Fig. 1**).

Immediately before each MRI examination, both superficial and deep temperature of muscles were measured at the mid thickness of the pig leg with a CE digital thermometer TP101 by using a sharp probe which temperature measure range from -50° to 300°. This was done in order to measure the decrease of the post-mortem muscle temperature (PMMT), which is a consecutive cadaveric phenomenon, linked to environmental and cadaveric factors. Measurements were performed close to the area included in the MRI scan (**Fig. 1**).

All MRI examinations were carried out with a 1.5 Tesla scanner (Ingenia, Philips, The Netherlands) using a 16-channel body coil.

Measurements of isotropic and anisotropic diffusion [19, 20], magnetization transfer ratio (MTR) [21], and tissue susceptibility [22] were obtained using sequences widely validated in clinical MRI, and correlated with PMMT and histological findings.

MRI examination protocol included:

1. T1-weighted Spin Echo sequence: TR 684 ms, TE 10 ms, NSA 1, matrix 308x183mm, slice thickness (SL) 10mm, scan duration 2m07s. This images were used as anatomical evaluation.
2. Magnetization transfer (MT) T1 sequence: magnetization transfer TR 684 ms, TE 10 ms, NSA 1, matrix 308x183, slice thickness (SL) 10 mm, acquisition time 2m07s.
3. Susceptibility weighted imaging (SWI): 4echoes, TR 51 ms, TE First 12 ms, , flip angle 20, echo spacing 10 ms, matrix 304x224. Scan duration 3 minutes.
4. Diffusion weighted imaging (DWI): TR 2230 ms TE 77 ms, matrix 124x103mm, SL 10mm, b factor 3 (0,500,1000), NSA 2, scan duration 1m13s, fat suppression SPIR.
5. Diffusion tensor imaging (DTI): TR 3979 ms, TE 88 ms, matrix 160x128mm, SL 2.2mm, directional resolution 32 scan duration 8m53s

Total acquisition time was 18 minutes.

The post-processing was performed with a complementary console (IntelliSpace[TM] portal 7.0) by a radiologist (MG) with a twenty year experience in MR imaging and ten year experience in MR molecular imaging [23–25].

The following molecular imaging values of the muscles were obtained from the analysis of the MRI sequences for each scan:

1) Apparent diffusion coefficient (ADC) from DWI was automatically calculated by the accessory and measured using ROIs.
2) Fractional anisotropy (FA) from DTI was automatically calculated by the accessory and measured using ROIs
3) Magnetization transfer ratio (MTR) was computed as T1s + MTs/T1s where T1s is the magnitude of tissue signal before the MT pulse and MTs is the signal after the MT pulse has been applied
4) SWI sequence

Calculation of all molecular imaging values was made using measurements acquired in the same regions of interest (ROIs) drawn on T1 weighted images and spread on the other sequences with copy and paste tools. The ROI was drawn including the largest part of the muscles and excluding large vessels, bones and subcutaneous tissue.

Color-coded ADC maps were also obtained.

Tractography of the muscles was carried out using the post processing program of the Philips complementary console.

Finally MinIP reconstructions of the SWI sequences were carried out in order to detect the presence of air within vessels and muscles and compared with standard anatomical T1 weighted images.

At 0, 12, 24 and 36 hours from the beginning of the MRI study, four histological samples were obtained in order to correlate MRI findings with histological post-mortem changes.

Samples were fixed in 10% neutral buffered formalin for 24 hours at room temperature. Thereafter, they were transversally cut and paraffin embedded. 4 μm thick sections were cut from each paraffin block and stained with haematoxylin and eosin stain for evaluation at light microscopy.

## STATISTICAL EVALUATION

Pearson test (carried out with MedCalc Statistical Software version 15.8 (MedCalc Software bvba, Ostend, Belgium; https://www.medcalc.org; 2015) was used for evaluating statistical correlation between post-mortem interval and MRI data.

## RESULTS

### Apparent Diffusion Coefficient

We found a highly significative statistical inverse correlation between ADC and PMI (significance level $P< 0.0001$, correlation coefficient -0.96, 95% confidence interval for r -0.99 to 0.85).

The data obtained from ADC map showed a progressive decrease of the ADC value from $1.55 \times 10^{-3}$ mm$^2$/sec at 0 hours up to $0.93 \times 10^{-3}$ mm$^2$/sec at 36 hours (**Fig. 2**). ADC decrease was 40% and it was well depicted in the color-coded maps (**Fig. 3**). It is noteworthy that the slope of the curve was not constant and was greater in the first 12 hours (22.6%) than in the following 24 hours (17.4%). This behavior correlates very well with histology since the first histological post-mortem sample (time 0) showed normal fibers and regular cellular structure.

In the sample taken at 12 hours, muscle fibers were swollen and some "giant" cells could be seen. Intercellular spaces were tight (**Fig. 4a**) explaining the reduction of water molecules movement and the strong increase of restriction in this phase. On the other hand, in the sample taken 36 hours, muscle fibers were shrunk and enlargement of intercellular spaces could be seen. Some fibers were missing and replaced by fluid-filled channels (**Fig. 4b**). Such a pattern of cellular degradation explain the lesser decrease of tissue restriction since in this phase reduction of water movement is due only to decrease of temperature.

**Magnetization Transfer Ratio**

We found a significative statistical inverse correlation between MTR and PMI (significance level P< 0.0012, correlation coefficient -0.86, 95% confidence interval for r: -0.96 to 0.52).

The values of MTR changed from 0.265 at 0 hour up to 0.133 at 36 hours with overall reduction of 49.8% (**Fig. 5**). In the curve 3 segments could be seen. The first segment between 0 and 12 hours was characterized by a constant decrease of the MTR values equal to 24,1%. In the second segment, between 12 and 28 hours, the decrease of MTR almost halted (from 24.1 up to 26.8%). Finally between 28 and 36 hours a rapid decrease of MTR values was detected (from 26.8% to 49.8 %).

**Fractional Anisotropy and Tractography**

We found a significative statistical inverse correlation between FA and PMI (significance level P< 0.0002, correlation coefficient -0.91, 95% confidence interval for r: -0.97 to 0.66).

FA dropped from 0.32 at 0 hour to 0.25 at 36 hours (21.9%).

The slope curve showed 3 segments (**Fig. 6**):

  1) First segment from 0 to 12 hours characterized by a decrease of 12.5%

  2) Second segment from 12 to 28 hours without variation

  3) Third segment from 28 to 36 hours with a further drop of 9.4% from 0.28 to 0.25

Tractography also demonstrated the decrease of FA as an evident loss of the number of muscular tracts (**Fig. 7 a and b**).

**Susceptibility**

SW images allowed detecting air in the vessels already in the first scan at 0 hour. Air was seen within the muscles after 12 hours (**Fig. 8**) but was never seen on standard T1 weighted images

**Temperature**

During the time of the experiment temperature of the phantom decreased from 35.7 up to 18.2 Celsius degrees (49%). Decrease of temperature was faster in the first 12 hours than in the following 24 hours. These data are in keeping with the post-mortem cooling of the human tissues [26].

## DISCUSSION

### Diffusion and apparent diffusion coefficient

Diffusion-weighted imaging (DWI) is a form of molecular MR imaging based on the random Brownian motion of water molecules within a voxel of tissue [19, 20, 27]. Diffusion is fundamentally a thermodynamic phenomenon as demonstrated by Einstein [28]. The state of extracellular space in tissues is the most important factor that regulates diffusion. Free movement of the water molecules in extracellular space is mainly restricted by the presence of cellular membranes. Highly cellular tissues or those with cellular swelling due to ischemia produce narrowing of the extracellular space and consequently exhibit lower diffusion movement, which can be measured by DWI, appearing as a reduction of ADC values. ADC is a quantitative parameter that is calculated from the DWI, it allows an objective evaluation of the water movement in tissue. For this reason diffusion MRI is particularly useful in diagnosis of malignant neoplasia and cerebral ischemia because of the high cellular tissue in malignant neoplasia and the cellular swelling due to anoxia in ischemia cause reduction of extracellular space and drop of ADC values.

The structure of muscle fibers generates anisotropy in the water diffusion in a similar way than axons in brain. So anisotropic diffusion represents a useful tool in evaluation of normal and diseased muscles [29, 30].

In our study we demonstrated that ADC values decreases of 40 % in 36 hours and correlates very well with PMI (Fig. 2 and 3).

Decrease of ADC values was greater in the first 12 hours (22.6%) with respect to the second 24 hours (17.4%). Comparison with histology allows us to explain this behavior since water restriction was influenced in the first post-mortem 12 hours by two synergic phenomena: 1) swelling of ischemic cells, and 2) decreasing temperature. On the other end 36 hours histology shows cells shrinkage and enlargement of the intercellular spaces which causes increasing of the water molecules movement, so in this phase reduction of ADC values is due only to decrease of temperature according to the Einstein – Stokes equation $D=kT/6phr$ where $k$ is the Boltzmann constant, $T$ is the absolute temperature in degrees Kelvin, $h$ is the viscosity of the medium, and $r$ is the radius of the molecule [28].

### Anisotropic Diffusion And Fractional Anisotropy

Anisotropic diffusion occurs in highly structured biological tissues that have different diffusion coefficients along different directions. White matter and muscles are highly anisotropic because of the parallel orientation of white matter and muscle fibers tracts [29]. The degree of anisotropic diffusion can be evaluated with diffusion tensor imaging (DTI) and it is well expressed by fractional anisotropy (FA), which is an index for the amount of asymmetrical diffusion within a voxel. The value of FA varies between 0 and 1. For perfect isotropic diffusion FA is 0. With progressive diffusion anisotropy, the FA $\rightarrow$ 1. Damage of the normal structure of an anisotropic diffusion can cause reduction of the FA with tendency of FA to move near 0.

Since DTI is sensitive to the orientation and density of cellular structures that hinder water diffusion, the local tissue microstructure can be evaluated with fractional anisotropy, which correlates with muscle fibers loss [30].

Anisotropic diffusion can be encoded to generate tractography images, which allow to represent white matter and muscle fibers as colored tracts. Therefore, loss of FA due to muscle damage causes changes in the tractography images. We observed a reduction of the FA up to 22%. This data correlates with histology showing gradual loss of muscle cells and enlargement of the intercellular space, that causes reduction of the muscle FA. Tractography demonstrates with better advantages the loss of muscle fibers (Fig. 4 and 7).

**MAGNETIZATION TRANSFER RATIO**

Magnetization transfer ratio (MTR) is an MRI technique that allow to enhance contrast between tissues where hydrogen protons are present in two states

1: bound to macromolecules

2: in free water

MT is based on the application of an off-resonance RF pulse, which saturates protons bound to macromolecules but not those in free water. The saturated protons from the macromolecule partially transfer their magnetization to protons in free water that in this way become partially saturated. When another radiofrequency pulse is applied, the signal from the free water is reduced due to the pre-saturation of this free-water protons. The difference between the signals achieved with and without the off-resonance pulse can be calculated and is referred as MTR [31].

MTR is a powerful tool in evaluating macromolecular compartment since damage of macromolecules causes decrease of the MTR. E.g., decrease in the magnetization transfer ratio have been shown to correlate well with the degree of myelin loss and axonal damage in patients with multiple sclerosis [32].

The MTR allows a quantitative evaluation of the ability of protons bound to macromolecules to exchange magnetization with the surrounding free water and, consequently, is able to detect damages of these macromolecules.

In our experiment MTR is reduced by half after 36 hours (Fig. 5). These data correlates well with alterations of the proteome profile which occurs few hours after death both in human and pig muscles. In samples of human muscles stored at 25° some proteins started to degrade after few hours [33].

Pittner et al. [9] showed that in pig muscles stored at 21°C proteins degraded in a regular and predictable fashion. Early change in the first 36h post mortem interval affected some proteins as Titina 1, desmin dp2, 1200 KDa and calpain 1.

The time-dependent changes in muscle histology are one of the useful indices for determining the postmortem interval. Vacuolization and autolysis occurred as early as 6 h when the muscle tissues were stored at 25°C [34, 35].Degradation of myofibrils by various proteases has been thought to be an important mechanism leading to postmortem vacuolization and autolysis in skeletal muscles [34, 35].

**SUSCEPTIBILITY**

Magnetic susceptibility corresponds to the internal magnetization of a tissue resulting from the interactions with an external magnetic field. Although human body is mainly diamagnetic, it contains paramagnetic substances as deoxy-hemoglobin. Moreover, tissue can contain pathological amount of paramagnetic substances e.g. blood degradation products as hemosiderin or ferritin. Finally, when two tissues with different magnetic susceptibilities are juxtaposed, local distortions in the magnetic field were seen. There are such natural interfaces between air and tissue.

Susceptibility weighted imaging (SWI) is an MRI Gradient Echo Steady State T2/T1 weighted sequence which is particularly sensitive to compounds that distort the local magnetic field [22, 36, 37].

The presence of paramagnetic substances or air in tissue causes large distortions in the magnetic field and significant susceptibility artifacts since the local field inhomogeneity accelerates transverse relaxation and signal decay (Fig. 8).

For this reason SWI is a tool with an extraordinary ability for detecting microscopic quantity of substances causing susceptibility artifact in tissues including air.

Presence of air in the vessels and tissue can be correlated to several conditions occurred before (e.g. trauma, infection, gas embolism) or after death (e.g. resuscitation procedures, putrefaction) and is considered a useful complementary tool in estimating post-mortem interval [15].

In our phantom SWI was able to detect air in vessels since the first scan due to exsanguination. Air in the vessels could be seen only in SWI but not in other sequences (Fig 8).We believe that SWI could become the method of choice for detecting micro-bubbles in tissues as an early marker of putrefaction. Further study with corpses in more advanced stage of decomposition are necessary to confirm this hypothesis.

It is necessary to point out the limitations of our study.

First, we obtained our data from a pig phantom. However literature data shows that pig's muscles are very similar to human muscles [8], so our data could be considered useful also for application in post-mortem qMMRI in humans.

Second, our data were collected only with phantom stored at 20° Celsius. Since the temperature storage is very important in causing post-mortem cellular changes, other data scanning muscles at different temperature storage should be acquired.

Third, a larger number of pig phantoms or humans should be acquired to amplify and confirm our results.

In spite of these drawbacks, we believe that our study demonstrates that multiparametric qMRMI is very sensitive to early post-mortem muscular changes due to cooling, damages of proteome probed by MTR and muscle fibers loss due to cell death showed by ADC and FA.

Consequently qMRMI has a potential for being a useful tool in forensic EPI delimitation with a high level of precision because the behavior of MRI signal is linked to the bio-physical post-mortem changes of tissues. This novel MRI approach also provides the following important advantages: it is a fast and relatively inexpensive method, and its results are not dependent from radiologist's subjective evaluation

In conclusion, qMRMI has the potential to evaluate EPI supporting other existing techniques. qMRMI could be included in a post-mortem multidisciplinary scenario in which, integrating data from different medical branches, a reliable (as it were "mathematical") estimation of PMI could be obtained [38].

*Compliance with Ethical Standards*
Funding: Not funding.
Conflict of Interest: the Authors confirm that there are no known conflicts of interest associated with this publication and there has been no financial support for this work that could have influenced its outcome.
Ethical approval: No permission by our Ethic Committee was necessary because the phantom was obtained from the commercial food chain.
Informed Consent: not needed.

**FIGURE CAPTIONS**

Figure 1. Picture of the pig leg. Black lines shows the area scanned with MRI. Arrows indicates the points used to carry out biopsies and temperature measurements.

Figure 2. a) Graphic of the changes of ADC. On the axis of abscissas are represented the time and the changes of temperature. b) Graphic of the correlation evaluated by Pearson's Test between ADC data and post-mortem timing

Figure 3. Panel of the color coded maps obtained at time 0 (upper map) and at 36 hours (lower map). Note the change of color from red to yellow-green representing decrease of the water diffusion in muscles.

Figure 4. a) Histological transverse section of pig muscle at 12 hours. Muscle fibers are swollen and surrounded by edematous spaces (Haematoxylin and eosin stain; original magnification, x100). b) Histological transverse section of pig muscle at 36 hours. Muscle fibers are shrunk or missing and replaced by fluid-filled channels (Haematoxylin and eosin stain; original magnification, x100).

Figure 5. a) Graphic of the changes of Magnetization Transfer Ratio. On the axis of abscissas are represented the time and the changes of temperature. b) Graphic of the correlation evaluated by Pearson's Test between Magnetization Transfer Ratio data and post-mortem timing

Figure 6. Graphic of the changes of Fractional Anisotropy. On the axis of abscissas are represented the time and the changes of temperature. b) Graphic of the correlation evaluated by Pearson's Test between Fractional Anisotropy data and post-mortem timing

Figure 7. a) Tractography obtained from diffusion tensor sequence at time 0. The fiber muscles are well depicted and appear quite compact b). Tractography obtained at 36 hours shows a dramatic loss of muscles fibers. The tractography correlates very well with histology shown in fig. 1b.

Fig. 8. Panel depicting the high sensititivity of susceptibility weighted imaging in demonstrating air in the vessels and tissue. Susceptibility weighted image shows with great advantage air in the vessels in the intermuscular vessels (arrows) and in the muscles (arrowheads). In T1 weighted image obtained in the same session (lower image) intravascular air cannot be seen.

**AUTHORS CONTRIBUTION**

- Daniela SAPIENZA, Michele GAETA, Alessio ASMUNDO: guarantors of integrity of entire study; study concepts and design;
- Daniela SAPIENZA, Salvatore SILIPIGNI, Ugo BARBARO, Antonella CINQUEGRANI, Patrizia GUALNIERA: literature research;
- Daniela SAPIENZA, Michele GAETA, Salvatore SILIPIGNI, Ugo BARBARO, Antonella CINQUEGRANI: data analysis/interpretation;
- Valeria BARRESI: histological data;
- Salvatore SILIPIGNI, Antonio BOTTARI: statistical analysis;
- Salvatore SILIPIGNI, Ugo BARBARO, Antonella CINQUEGRANI, Francesca GRANATA: manuscript preparation;
- ALL AUTHORS: manuscript final version approval.

The Authors confirm that there are no known conflicts of interest associated with this publication and there has been no significant financial support for this work that could have influenced its outcome.

All authors read and approved the final manuscript and they confirm that there are no other persons who satisfied the criteria for authorship but are not listed. The Authors further confirm that the order of authors listed in the manuscript has been approved by all of us.